\documentclass[journal,twoside,web]{IEEEtran}

\usepackage{cite}
\usepackage{amsmath,amssymb,amsfonts}
\usepackage{algorithmic}
\usepackage{graphicx}
\usepackage{textcomp}
\usepackage{hyperref}

\usepackage{textcomp}
\usepackage{wrapfig}
\usepackage{subfigure}
\usepackage{tabulary}
\newcolumntype{K}[1]{>{\centering\arraybackslash}p{#1}}
\usepackage{booktabs}
\usepackage{multirow}
\usepackage{subfloat}
\usepackage{float}
\usepackage{adjustbox}
\usepackage{caption}
\usepackage{soul}

\def\BibTeX{{\rm B\kern-.05em{\sc i\kern-.025em b}\kern-.08em
    T\kern-.1667em\lower.7ex\hbox{E}\kern-.125emX}}
\begin{document}
\title{ANet: Autoencoder-Based Local Field Potential Feature Extractor for Evaluating
An Antidepressant Effect in Mice after Administering Kratom Leaf Extracts}
\author{Jakkrit~Nukitram, 
        Rattanaphon~Chaisaen,
        Phairot~Autthasan,
        Narumon~Sengnon,
        Juraithip~Wungsintaweekul,
        Wanumaidah~Saengmolee,
        Dania~Cheaha,
        Ekkasit~Kumarnsit,
        Thapanun~Sudhawiyangkul$^{*}$,
        and Theerawit~Wilaiprasitporn$^{*}$ \IEEEmembership{Senior Member, IEEE}
\thanks{This work was financially supported by the professional development project under the Science Achievement Scholarship of Thailand (SAST). A partial grant was also offered by the Program of Physiology, Division of Health and Applied Science, Faculty of Science, Prince of Songkla University, Thailand.}
\thanks{J. Nukitram and E. Kimarnsit are with the Physiology Program, Division of Health and Applied Sciences, and Biosignal Research Center for Health, Faculty of Science, Prince of Songkla University (PSU), Hatyai Campus, Hatyai, Songkhla 90112, Thailand.}
\thanks{D. Cheaha is with the Biology Program, Division of Biological Science, and Biosignal Research Center for Health, Faculty of Science, Prince of Songkla University, Hatyai, Songkhla, Thailand.}
\thanks{N. Sengnon and J. Wungsintaweekul are with Department of Pharmacognosy and Pharmaceutical Botany, Faculty of Pharmaceutical Sciences, Prince of Songkla University (PSU), Hatyai Campus, Hatyai, Songkhla 90112, Thailand.}
\thanks{W. Saengmolee, R. Chaisaen, P. Autthasan, T. Sudhawiyangkul and T. Wilaiprasitporn are with  Bio-inspired Robotics and Neural Engineering (BRAIN) Lab, School of Information Science and Technology (IST), Vidyasirimedhi Institute of Science \& Technology (VISTEC), Rayong 21210, Thailand (e-mail: Thapanun.s@vistec.ac.th; Theerawit.w@vistec.ac.th)\newline 
$^{*}$\textit{Corresponding author: Thapanun Sudhawiyangkul and Theerawit Wilaiprasitporn}}
}

\maketitle

\begin{abstract}
Kratom (KT) typically exerts antidepressant (AD) effects. However, evaluating which form of KT extracts possesses AD properties similar to the standard AD fluoxetine (flu) remained challenging. Here, we adopted an autoencoder (AE)-based anomaly detector called \textit{ANet} to measure the similarity of mice's local field potential (LFP) features that responded to KT leave extracts and AD flu. The features that responded to KT syrup had the highest similarity to those that responded to the AD flu at 85.62 ± 0.29\%. This finding presents the higher feasibility of using KT syrup as an alternative substance for depressant therapy than KT alkaloids and KT aqueous, which are the other candidates in this study. Apart from the similarity measurement, we utilized ANet as a multi-task AE and evaluated the performance in discriminating multi-class LFP responses corresponding to the effect of different KT extracts and AD flu simultaneously.
Furthermore, we visualized learned latent features among LFP responses qualitatively and quantitatively as \textit{t}-SNE projection and maximum mean discrepancy distance, respectively. The classification results reported the accuracy and F1-score of 79.78 ± 0.39\% and 79.53 ± 0.00\%. In summary, the outcomes of this research might help therapeutic design devices for an alternative substance profile evaluation, such as Kratom-based form in real-world applications.
\end{abstract}

 


\section{Introduction}
\label{sec:introduction}
\IEEEPARstart {K}{ratom} is a local name of \textit{Mitragyna speciosa} (Korth.) Havil. and has been broadly recognized as herbal medicine for many decades to treat and improve withdrawal symptoms effectively and psychological disorders induced by abused drugs in animal models \cite{kumarnsit2007fos} and humans \cite{assanangkornchai2007use}. On a pre-clinical scale, many forms of KT leave extracts have been reported to possess antidepressant (AD)-liked activity on the central nervous system (CNS) \cite{cheaha2015effects,kumarnsit2007effect}. As a result, it was still unclear which kind of KT leaf extract might have a profile similar to standard AD.

Local field potential (LFP) recording is an invasive measurement to capture brain working, providing high temporal resolution and sensitivity. The spectral power of LFP oscillates dependently on the alteration of the neuro-biomolecules (NeuroBios) and is a pivotal character in drug profile classification in rodents \cite{manor2021characterization}. However, the only statistical analysis may face errors when there are substantial medical databases to be distinguished, and each has similar effects on LFP features. Therefore, the machine learning approach is an alternative way for users to recognize LFP features in higher complexity tasks \cite{korotcov2017comparison}. In the literature, only a few experiments attempted extracting LFP essential features from animals with the deep neural network, which is similar to our interest. However, the publication above primarily relied on a simple convolutional neural network \cite{ali2022predictive}, which may not be a complete pipeline distinctively. Therefore, according to our motivation, we desire to propose a novel pipeline for LFP processing.

A compressed sensing method, autoencoder (AE), is a computational-based deep learning architecture or model \cite{gogna2016semi}. AE primarily consists of two components: an encoder and a decoder. Mapping the input signals to latent space is the encoder network's main action; meanwhile, another network's function is reconstructing the latent vector to be the input signals at the end of the model \cite{vincent2010stacked}. Integrating a supervised classifier network to the AE, so-called \textit{multi-task AE}, effectively learning to compress data, reconstruct, and classify the brain signals simultaneously \cite{ditthapron2019universal}. In terms of application, although tiny alteration of the brain working in a fraction of a millisecond, such as during performing imaginary motor tasks as well as examining the test to diagnose developmental dyslexia, which is generally difficult to distinguish by visual inspection, the multi-task AE can classify well by having the brain signals or EEG as the inputs to the models. \cite{autthasan2021min2net,martinez2020eeg}. Meanwhile, as far as we know, none of the previous works apply the multi-task AE to process the LFP signals. Thus the study related to mice's LFP response to given substance, as in the rest of the paper, is a novel issue.

\begin{figure*}
\centering
\includegraphics[width=17 cm,height =6.5 cm]{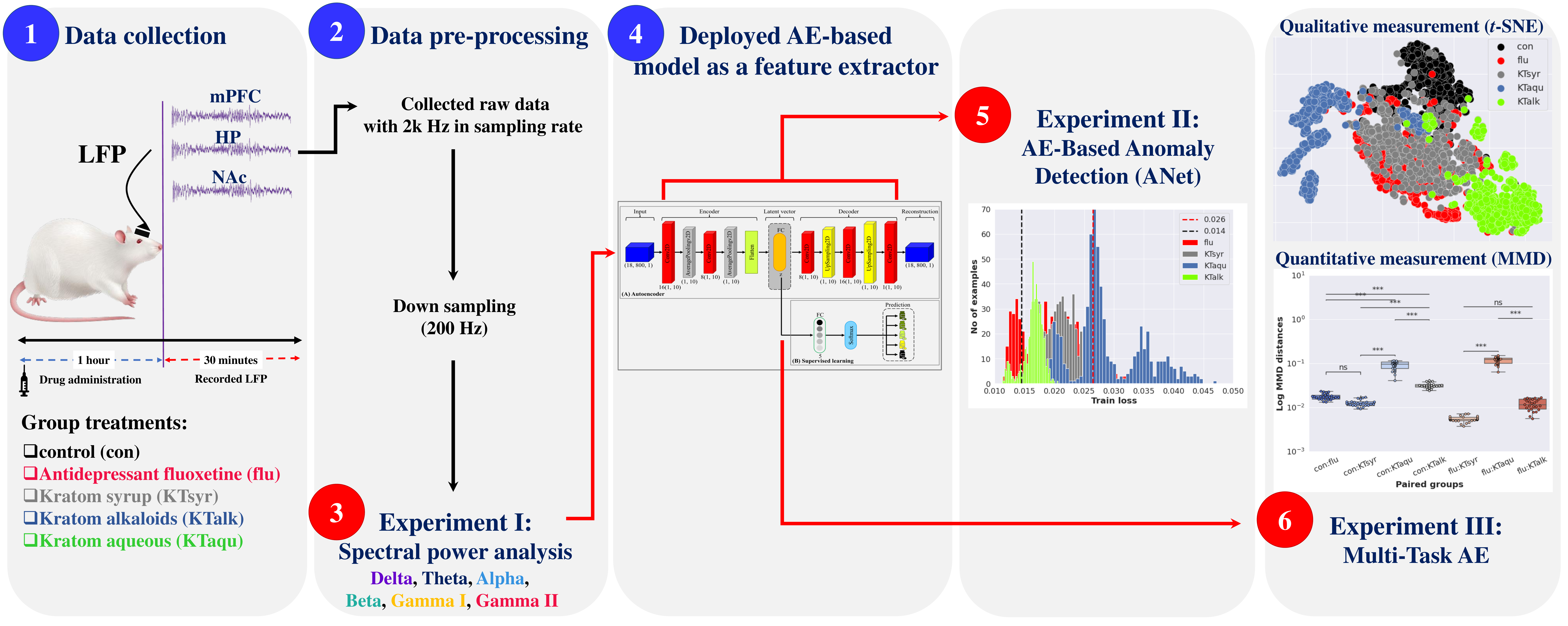}
\caption{An overview of this study. \textbf{Note:} Medial prefrontal cortex (mPFC), Hippocampus (HP), Nucleus accumbens (NAc), control (con), fluoxetine (flu), KT syrup (KTsyr), KT alkaloids (KTalk), and KT aqueous (KTaqu)}
\label{fig:Fig1}
\end{figure*}

Here, we proposed the AE-based anomaly detector called \textit{ANet} as the LFP feature extractor that can measure the similarity of mice's LFP administering KT extracts and AD fluoxetine (flu). \autoref{fig:Fig1} illustrated ANet (AE) and extended ANet (multi-task AE, which is AE and supervised learning) architecture or model with preprocessed LFP as the input to the model.
\begin{itemize}
\item ANet is the novel mice's LFP feature extractor automatically extracts the discriminative features from LFP in responses to treated substances. These features play significant roles in the LFP similarity measure or comparison in the form of anomaly detection compared to the target LFP, which responds to this study's standard drug, AD flu.

\item We extended ANet to be the multi-task AE in the feasibility study of simultaneous classifying multiple LFP responses induced by the multiple KT extracts and AD flu. Apart from classification performance, we evaluate the feature extraction capability through the latent space qualitatively and quantitatively as \textit{t}-distributed stochastic neighbor embedding (\textit{t}-SNE) projection and maximum mean discrepancy (MMD) distance, respectively.

\item Based on mice's brain responses or LFP evaluated using ANet and the extended version (the multi-task AE), we found that KT syrup produced the highest similar LFP features to AD flu. According to the literature and the proposed findings, we are the first to reveal that using KT syrup extract is an appropriate candidate substance for depressant therapy. However, further confirmation of the clinical study on AD effects from KT syrup remains a future challenge.
\end{itemize}

\section{Materials and Data Gathering}
\subsection{Kratom Leave Extracts}
We collected KT leave from the natural resources in Surat Thani province, Thailand. Supplementary materials explain more details about the preparation of KT extract and the quantification of KT major alkaloids, mitragynine (MT), that accumulated in each form of KT extracts.
\subsection{Mice}
The committee approved all operations involving animals employed in the scientific study from the Prince of Songkla University's Institute of Animal Care and Use, which followed the criteria of the International Committee on Laboratory Animal Science (ICLAS) [project license number: MHESI 6800.11/845 and reference number: 57/2019]. We recruited Male Swiss albino ICR mice from the Nomura Siam International Company, Bangkok, Thailand. They were appropriately at rest for one week before the experiment began to minimize the stress. Moreover, mice stayed in separated stainless steel cages (17 x 28.5 x 17 cm) under the standard condition room (12/12 h light/dark cycle, 22°C, and 55.1$\%$ relative humidity). Water and commercial food pellets were accessed freely. We conducted the studies between eight a.m. and four p.m.

\subsection{LFP Electrode Implantation}
Male ICR mice (four months of age) had been through an intraperitoneal injection of a mixed solution of 16 mg/kg xylazine hydrochloride (Xylavet, Sigma-Aldrich International GmbH, Switzerland) and 50 mg/kg zoletil (Tiletamine – zolazepam, Vibac Ah, Inc., USA) to be deeply anesthetized. The head of the animals was then fixed with stereotaxic apparatus before the scalp on the dorsal head in the middle line was exposed. According to the mouse brain atlas \cite{paxinos2019paxinos}, electrodes were stereotaxically implanted on the left hemisphere of the brain, from the bregma to the medial prefrontal cortex (mPFC) (AP; +2.5 mm, ML; 0.2 mm; DV; 1.5 mm), hippocampus (HP) (AP; -2.5 mm, ML; 2.0 mm; DV; 1.5 mm) and the nucleus accumbens (NAc) (AP; +1.3 mm, ML; 1.0 mm; DV; 4.2 mm). Over the cerebellum (AP; -6.0 mm, ML; 0.0 mm; DV; 1.5 mm), a ground electrode was inserted. Dental acrylic was applied to hold and secure all placed electrodes. The antibiotic ampicillin (100 mg/kg) (General Drug House Co., Ltd., Bangkok, Thailand) and carprofen (10 mg/kg) (Best Equipment Center Co., Ltd., Thailand) were given intramuscularly once a day for three days \cite{nukitram2022ameliorative} to prevent infection and relieve pain. It took at least two weeks to fully recuperate from surgery before they could begin the experiment.

\subsection{Data Collecting in Mice}
As soon as animals recovered from surgery, they went to the experimental period adapted from the previous investigations \cite{cheaha2017effects,manor2021characterization}, as shown in supplementary materials. In brief, after animals habituated for three consecutive days, they went to the testing phase. During this phase, we placed mice individually in the recording chamber (25 cm x 18 cm x 25 cm) to familiarize the experimental conditions for one hour. After that, the LFP signals were harvested in the recording session for 30 minutes after the injection of the treated substances (number of mice = 7 per group) for one hour. The concentrations of KT extract in each sample existed in \autoref{tab:Table1} according to the high-performance liquid chromatography analysis. KT extract calculated doses are from the whole amount of sample containing MT, fixed at 10 mg/kg, a dose that effectively cured the animal model of depression \cite{idayu2011antidepressant}. The AD flu at 10 mg/kg was selected as the standard drug since it showed positive results in previous findings \cite{cheaha2015effects}.

Details of instrumental setup for LFP signals recording are with the previous works \cite{nukitram2022ameliorative}. In summary, the Dual Bio Amp (AD Instruments, Castle Hill. NSW, Australia) and the PowerLab 16/35 system (AD Instruments, Castle Hill. NSW, Australia) with 16-bit A/D at a sampling frequency for two kHz are amplification and digitization for LFP signals, respectively. Also, we filtered the power line noise artifact at 50 Hz. LabChart 7.3.7 Pro software was a tool for recording LFP signals.

\begin{table} 
\caption{Summary of MT content in KT extracts used in this study}
\label{tab:Table1}
\resizebox{\columnwidth}{!}{%
    \LARGE
\begin{tabular}{l c c c} 
    \toprule[0.2em]
    \multicolumn{1}{c}{\textbf{Sample}} 
    & \textbf{\begin{tabular}[c]{@{}c@{}}MT content \\(Mean ± SD)\end{tabular}}
    
    & \textbf{\begin{tabular}[c]{@{}c@{}}Amount of MT per \\ administration \end{tabular}}
    
    & \textbf{\begin{tabular}[c]{@{}c@{}}Dose of each sample \\ per administration \end{tabular}}\\ \midrule[0.1em]
    KT alkaloids &47.10 ± 0.55 mg/g DW &10 &212 mg  \\ \midrule[0.1em]
    KT aqueous &10.80 ± 0.00 mg/g DW &10 &926 mg  \\ \midrule[0.1em]
    KT syrup &2.30 ± 0.00 mg/mL &10  &4.35 mL \\
    \bottomrule[0.2em]\\

    \multicolumn{4}{l}{\textbf{Note:} DW designated as dry weight. The samples were analyzed in triplicate.}
\end{tabular}

}
\end{table}

\section{Experiments}
This work has designed three experiment settings: Spectral Power with Statistical Analysis, AE-Based Anomaly Detection, and Multi-Task AE. The purpose and detail of each experiment are described below.

\subsection{Experiment I: Spectral Power with Statistical Analysis}
This experiment's purpose was to compare LFP responses induced by different substances traditionally. The substances used are control, flu, KT syrup, KT alkaloids, or KT aqueous. First, we downsampled raw LFP to 200 Hz using the MNE python package (version 0.23.4) \cite{gramfort2013meg}. Then we computed the spectral power of the prepared LFP for all channels by applying the Morlet wavelets transform and divided into six frequency bands: Delta, Theta, Alpha, Beta, Gamma I, and Gamma II according to the previous work \cite{nukitram2022ameliorative}. Finally, we conducted the statistical analysis using a non-parametric Kruskal-Wallis test with Bonferroni multiple comparisons to test group differences across the frequency bands at each brain region.

Comparing statistical results to the deep learning approaches, the recorded LFP signals at each brain region were further split into four seconds/segment, followed by spectral power of the frequency band extraction as the input to ANet and the multi-task AE in the following two experiments as shown in \autoref{fig:Fig2} and \autoref{fig:Fig3}.

\subsection{Experiment II: AE-Based Anomaly Detection (ANet)}
This experiment aimed to evaluate the similarity of LFP features responded to each KT extract relative to the reference drug (AD flu). We first designed the structure of the auto-encoder or ANet as the following two components.
\subsubsection{Encoder}
The encoder network contained two convolutional neural networks (CNN) stacks; each consisted of a 2-D convolutional layer (Conv2D) performed with an exponential linear unit (ELU) activation function, a batch normalization layer, and an averaging pooling layer (AveragePooling2D), respectively. In this module, the input format was (\textit{F}*\textit{B}, \textit{T}, 1), where \textit{F}, \textit{B}, and \textit{T} denote the number of frequency bands, brain areas, and time points, respectively. During the implementation, we set a data format as the \textit{channel last} option for Conv2D (Keras API). Flatten was the final layer before mapping to the latent representation or vector.
\subsubsection{Decoder}
The decoder aligned symmetrically with the encoder component to shape the dimension of reconstructed signals to be the exact size of the original input. Three CNN blocks reconstruct the information from the latent vector. Each CNN block of the decoder network consisted of a Conv2D activated by the ELU function, followed by a UpSampling2D layer to expand the spatial dimension of the data.

\begin{figure}
\centering
\includegraphics[width=8.5 cm,height =6.5 cm]{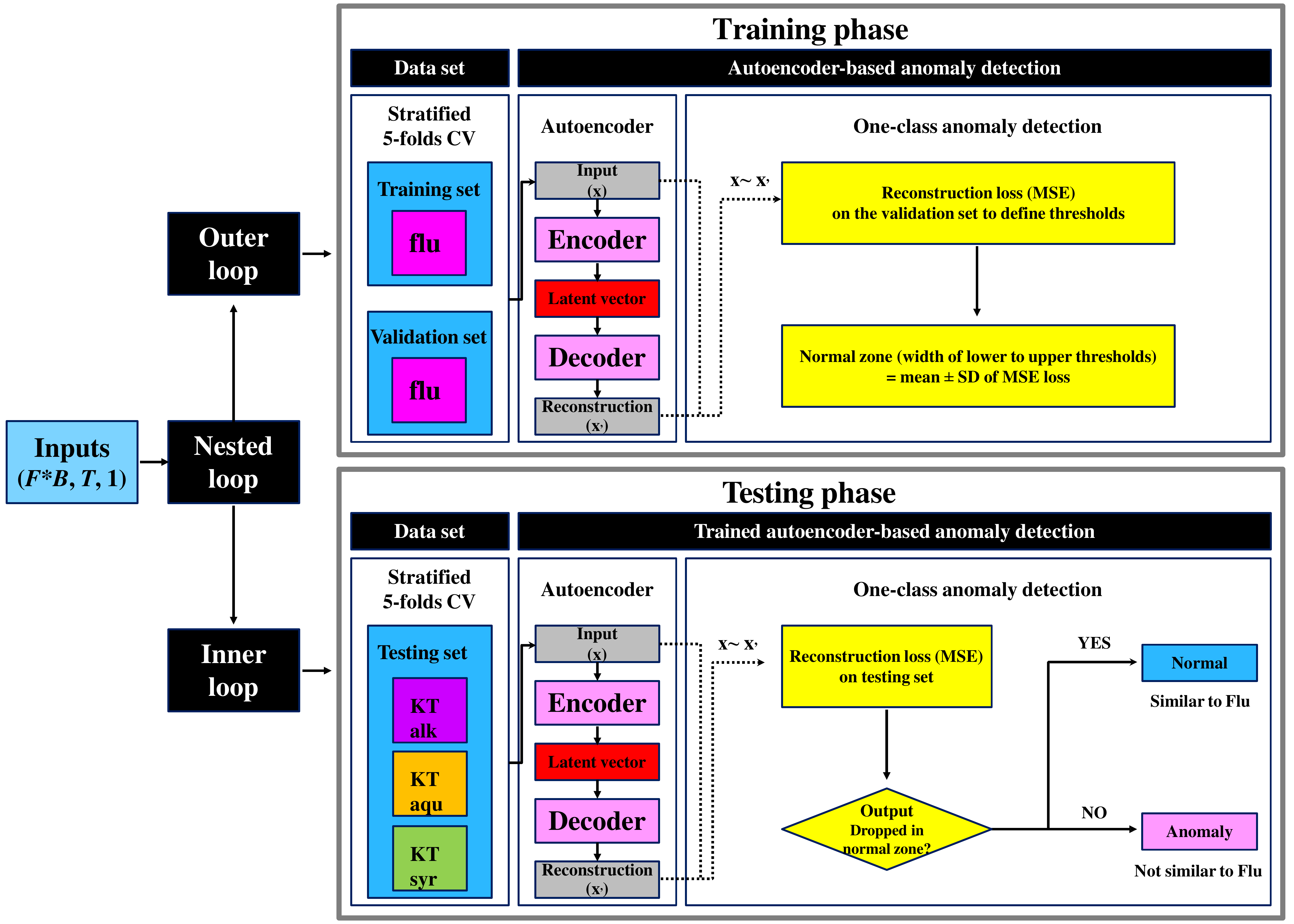}
\caption{The procedure of Experiment II: AE-Based Anomaly Detection (ANet) \textbf{Note:} Number of frequency bands (\textit{F}), Number of brain areas (\textit{B}), Number of time points (\textit{T}), the LFP features in response to each substance: KT syrup (KTsyr), KT aqueous (KTaqu), KT alkaloids (KTalk), and fluoxetine (flu)}
\label{fig:Fig2}
\end{figure}

To build ANet to be the anomaly detector, as illustrated in \autoref{fig:Fig2}, LFP spectral power responded to AD flu was the original input during training as the conventional AE. Therefore, we calculated mean square error (MSE) loss (Equation \eqref{MSE}) during ANet learning. The mean and standard deviation (SD) of MSE loss defined the threshold for the normal zone of the anomaly detector. The width of the normal zone was the mean ± SD. Then we used the trained ANet to evaluate the similarity of the unseen LFP features that responded to KT extracts (testing sets). If MSE loss of the unseen LFP was fallen into the normal zone, LFP features that responded to the KT extract are deemed similar to standard AD drugs and vice versa. Then we optimized the model according to the following Subsections \ref{Network training} and \ref{sssec:NetworkValidation}.

\begin{figure}
\centering
\includegraphics[width=8.5 cm,height =6.5 cm]{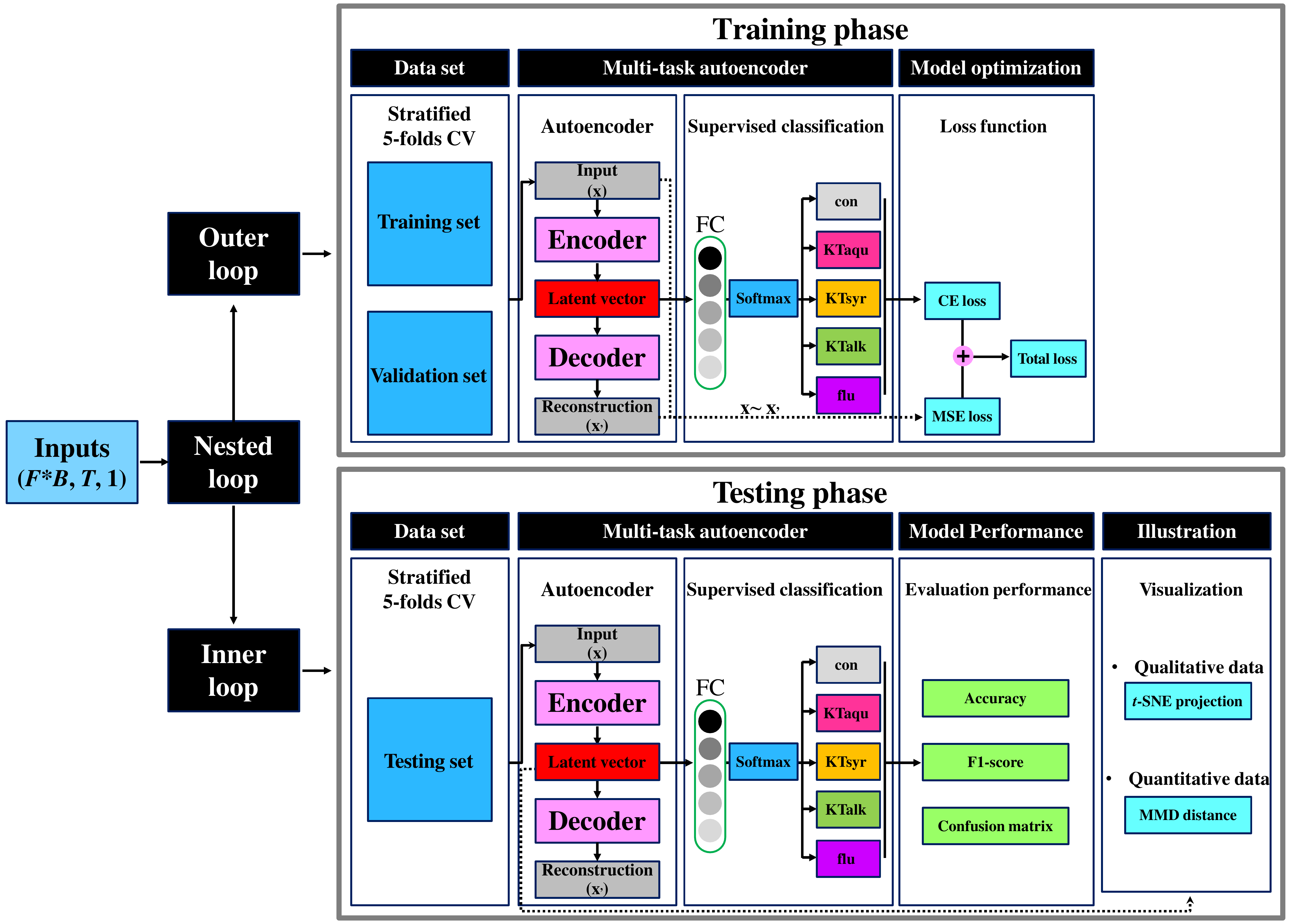}
\caption{The procedure of Experiment III: Multi-Task AE \textbf{Note:} Number of frequency bands (\textit{F}), Number of brain areas (\textit{B}), Number of time points (\textit{T}), the LFP features in response to each substance: control (con), KT syrup (KTsyr), KT aqueous (KTaqu), KT alkaloids (KTalk), and fluoxetine (flu)}
\label{fig:Fig3}
\end{figure}

\subsection{Experiment III: Multi-Task AE}
This study extended ANet by including the supervised learning component to the AE, which turned out to be the multi-task AE. We aimed to perform the feasibility of extracting discriminative features from multiple LFP responses simultaneously.

\textit{ANet and Supervised Learning:}
We added the supervised learning component by having the latent vector layer of ANet as the input, which made the proposed architecture form the multi-task AE, as shown in \autoref{fig:Fig1}. At the end of the supervised task, the softmax function classified the latent vector via layer activation in the FC layer. These allowed our model to learn the reconstruction along with the classification.

During the experiment, we explored the possibility of the multi-task AE or the model recognizing the multi-class LFP features for classifying LFP responses. We believed that this experiment could benefit in developing computational tools for future studies of the substance that affected mice's brain activities. The experimental protocol followed \autoref{fig:Fig3}. Firstly, we split the inputs for training, validating, and testing sets. Then we optimized the model according to the following Subsections \ref{Network training} and \ref{sssec:NetworkValidation}. Eventually, the following bullets and equation explain quantitative and qualitative assessments of the model capability in LFP feature extraction.

\begin{itemize}
\item \textit{$t$-SNE projection} is for the qualitative assessment of the  visualize reduced dimensional data \cite{van2008visualizing}. In this experiment, we defined the latent vector as the input of \textit{t}-SNE projection algorithm to present the probability distribution of unlearned and learned features, representing the LFP datasets before and after feature extraction by the proposed model, respectively.
  
\item \textit{Maximum Mean Discrepancy (MMD) distance} is for the quantitative assessment of the distance between the means of the embedding features of two probability distributions in \textit{t}-SNE projection. They evaluated whether data in set \textit{X} and \textit{Y} generated a probability distribution equally. Here, the ideal concept is that if the MMD distance of LFP features between substance \textit{X} and \textit{Y} is low. Then, we might imply that those two LFP inputs seem to originate from a very close origination. The standard calculation of MMD is the statistical measure of the samples from the different sets, expressed in Equation \eqref{MMD}.
\end{itemize}

\begin{equation}\label{MMD}
\small
M M D\left(P_{X}, P_{Y}\right)=\left\|\mathbb{E}_{X \sim P_{X}}[\varphi(X)]-\mathrm{E}_{Y \sim P_{Y}}[\varphi(Y)]\right\| \kappa
\end{equation}

Where $\varphi$ represents the map of features of $X$ or  $Y \rightarrow \kappa$, $\kappa$ denotes a reproducing kernel Hilbert space. $P_{X}$ and $P_{Y}$ represent the probability distribution of LFP in sets $X$ and $Y$, respectively. In addition, the multi-task AE performance also exhibited averaged percentage accuracy, F1-score, and confusion matrix.

\subsection{Network Training}\label{Network training}
In both Experiment II and III, we implemented our model using Keras API (TensorFlow version 2.2.0 as backend) under NVIDIA Tesla v100 GPU setup with 32G memory. With the optimization by Adam optimizer, the learning rate was between 10\textsuperscript{-4} and 10\textsuperscript{-5}. If there were no improvement in validation loss for five consecutive epochs, the learning rate gradually alleviated with a decay rate of 0.5. The batch size was 128 samples. We applied the function of early stopping to stop the training iteration when the validation loss had not reduced for 20 continuous epochs. 

According to the described experiments earlier, we optimized mean square error (MSE) loss alone for Experiment II or the study of ANet. In contrast, we optimized both MSE and cross-entropy (CE) losses for Experiment III, which studied the multi-task AE. MSE and CE losses correspond to AE and supervised learning components, respectively, as described below:

\begin{itemize}
    \item Mean square error (MSE) loss was adopted to monitor and minimize errors between the input signal and reconstructed data. It was expressed in Equation \eqref{MSE}.
\end{itemize}

\begin{equation}\label{MSE}
L_{M S E}(x, \hat{x})=\frac{1}{C} \sum_{i=1}^{C}\left\|x_{i}-\hat{x}_{i}\right\|^{2}
\end{equation}

$C$ represents a number of channels while $x$ and $\hat{x}$ are input and reconstructed data of the channel.\newline

\begin{itemize}
    \item Cross-entropy (CE) loss was used in supervised learning to assess the error between the actual label and classification probability, as shown in Equation \eqref{CE}.
\end{itemize}

\begin{equation}\label{CE}
L_{C E}(y, \hat{y})=-\sum_{j=1}^{|c|} y_{j} \log \hat{y}_{j}
\end{equation}

Where $y_{j}$ and $\hat{y}_{j}$ denote true and predicted labels of data in class, and $c$ is the number of classes.

The summation of the losses, as mentioned earlier, finally evaluated the total loss: $L_{M S E}$ and $L_{C E}$ demonstrated in Equation \eqref{MSE} and Equation \eqref{CE}, respectively, as shown below. 

\begin{equation}\label{total_loss}
L_{total}(x,\hat{x},y,\hat{y})= \frac{1}{N} \sum_{k=1}^{N}(L_{MSE}(x_{k},\hat{x}_{k}) + L_{CE}(y_{k},\hat{y}_{k}))
\end{equation}

$N$ is the total number of input signals and the weight of loss assigned for 1.0.

\subsection{Network Validation}
\label{sssec:NetworkValidation}
Experiment II validated ANet for anomaly detection with the subject-independent scheme. The stratified 5-folds cross-validation was adopted to separate data for testing, and training sets, both outer and inner loops, using Scikit-Learn (version 1.1.1). Only the spectral power of LFP that responded to the AD flu was assigned as training and validation sets, while unseen data from KT extracts were used as a testing set, as illustrated in \autoref{fig:Fig2}. 

The multiple LFP responses classification required a larger scale of samples in the network training, so Experiment III  validated the multi-task AE with the subject-dependent scheme. The spectral powers of multi-LFP classes responded to each substance: control, AD flu, KT syrup, KT alkaloids, and KT aqueous gathered from seven mice per class, as shown in \autoref{fig:Fig3}.

\section{Results}
\begin{table}
\centering
\Large
\caption{Summary of pairwise multiple comparison test of spectral power in selected paired substances. }
\label{tab:Table3}
\resizebox{1\columnwidth}{!}{%
    
\begin{tabular}{@{}ccccccccc@{}}
    \toprule[0.2em]
    \multirow{2}{*}{\textbf{Brain regions}} & \multirow{2}{*}{\textbf{Bands}} & \multicolumn{7}{c}{\textbf{Paired groups}}
    
    \\ \cmidrule[0.1em](l){3-9} 
    &   &con-flu  &con-KTsyr  &con-KTaqu  &con-KTalk  &flu-KTsyr  &flu-KTaqu  &flu-KTalk \\ \midrule[0.1em]

    \multirow{6}{*}{NAc} 
    
    &Delta  &    &     &      &      &      &       &  \\
    &Theta  &    &     &      &      &      &       &  \\
    &Alpha  &    &     &      &$*$      &      &       &$*$  \\
    &Beta   &    &$*$      &      &$*$      &      &       & \\
    &Gamma I   &    &      &      &      &      &       & \\
    &Gamma II   &    &      &      &      &      &       & \\ \midrule[0.1em]

    \multirow{6}{*}{mPFC} 
    
    &Delta  &    &     &      &$*$      &      &       &  \\
    &Theta  &$***$     &$***$     &      &$***$       &      &       &  \\
    &Alpha  &$***$     &$***$      &      &$***$       &      &$***$        &  \\
    &Beta   &$***$    &$***$     &      &$***$      &      &$**$       & \\
    &Gamma I   &$*$    &      &      &$**$      &      &$*$       & \\
    &Gamma II   &    &      &$***$      &      &      &$*$       &$*$ \\ \midrule[0.1em]

    \multirow{6}{*}{HP}
    &Delta   &  &   &   &   &   &   &\\
    &Theta   &  &   &   &   &   &   &\\
    &Alpha   &  &   &   &   &   &   &\\
    &Beta    &  &   &$*$    &   &   &$*$    &\\
    &Gamma I    &  &   &$***$   &   &   &$***$   &\\
    &Gamma II   &  &   &$***$   &   &   &$***$   &\\ 
                                      
    \bottomrule[0.2em]\\
    
    \multicolumn{9}{l}{\textbf{Note:} $*$, $**$ and $***$ represent  $p$ $<$ $0.05$, $0.01$, $0.001$, respectively}
    
 
\end{tabular}
    }
\end{table}

\subsection{Experiment I: Spectral Power with Statistical Analysis}\label{Experiment_I}
The conventional hand-crafted feature of LFP signals is spectral power. This experiment used statistical testing across the quantitative spectral power analysis to assess the significant difference in LFP responses among the group of mice treated with different substances. As reported in \autoref{tab:Table3}, the results found that LFP responses to KT syrup and KT alkaloids have no significant difference from AD flu in most brain regions and oscillation bands. In summary, the results convinced both KT extracts could be potential candidates for alternative drugs given the similar effects on the brain to AD flu. However, this conventional approach demonstrated the weakness in differentiating LFP responses from KT syrup and KT alkaloids. Thus, we addressed the weakness by proposing ANet and the multi-task AE approaches in identifying the higher similarity of LFP responses to KT syrup and AD flu compared to those to KT alkaloids.

\begin{figure}
    \centering
    \subfigure []{
        \label{subfig:Fig9A}
        \includegraphics[width=0.225\textwidth]{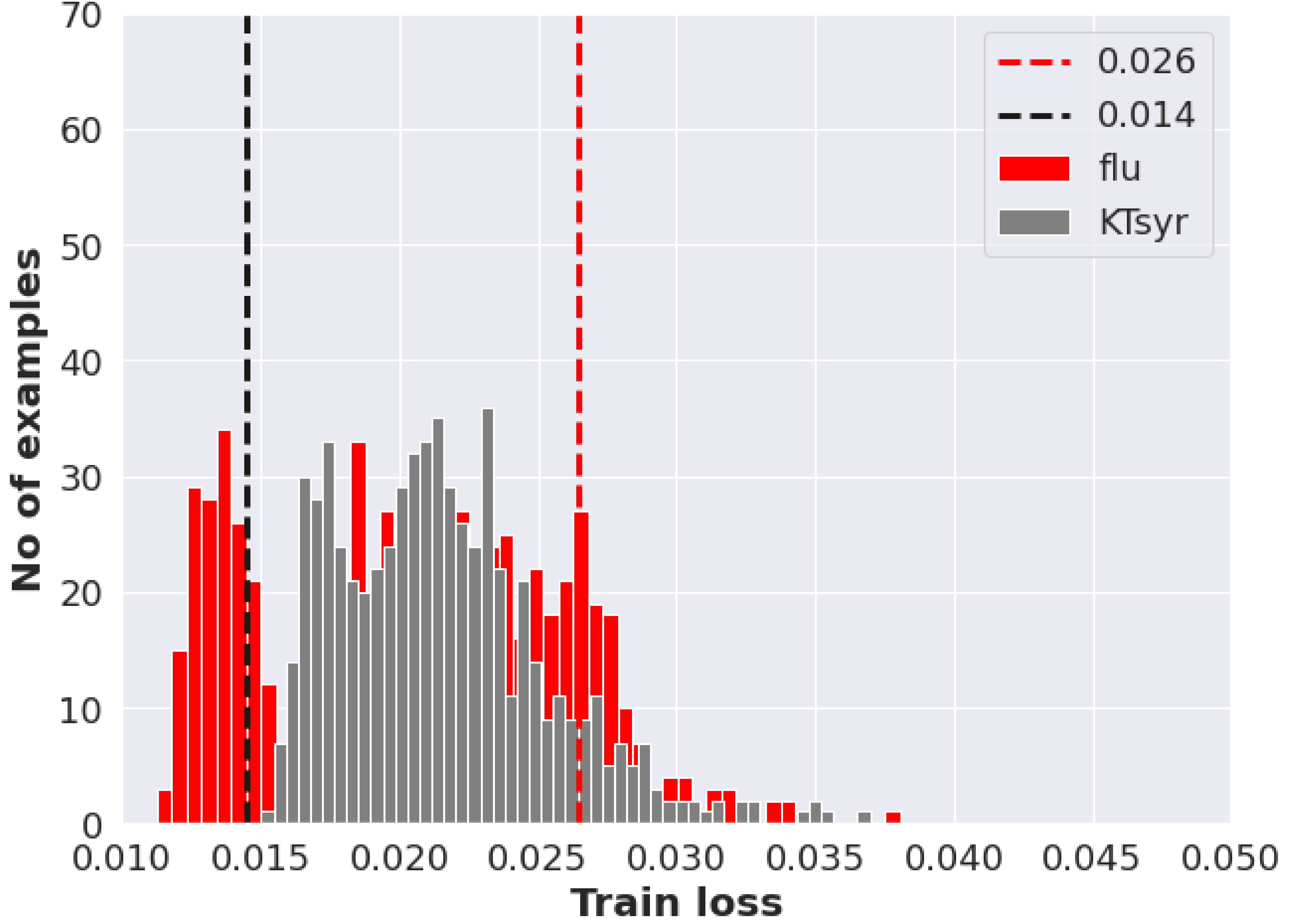} 
    }
    \subfigure []{
        \label{subfig:Fig9B}
        \includegraphics[width=0.225\textwidth]{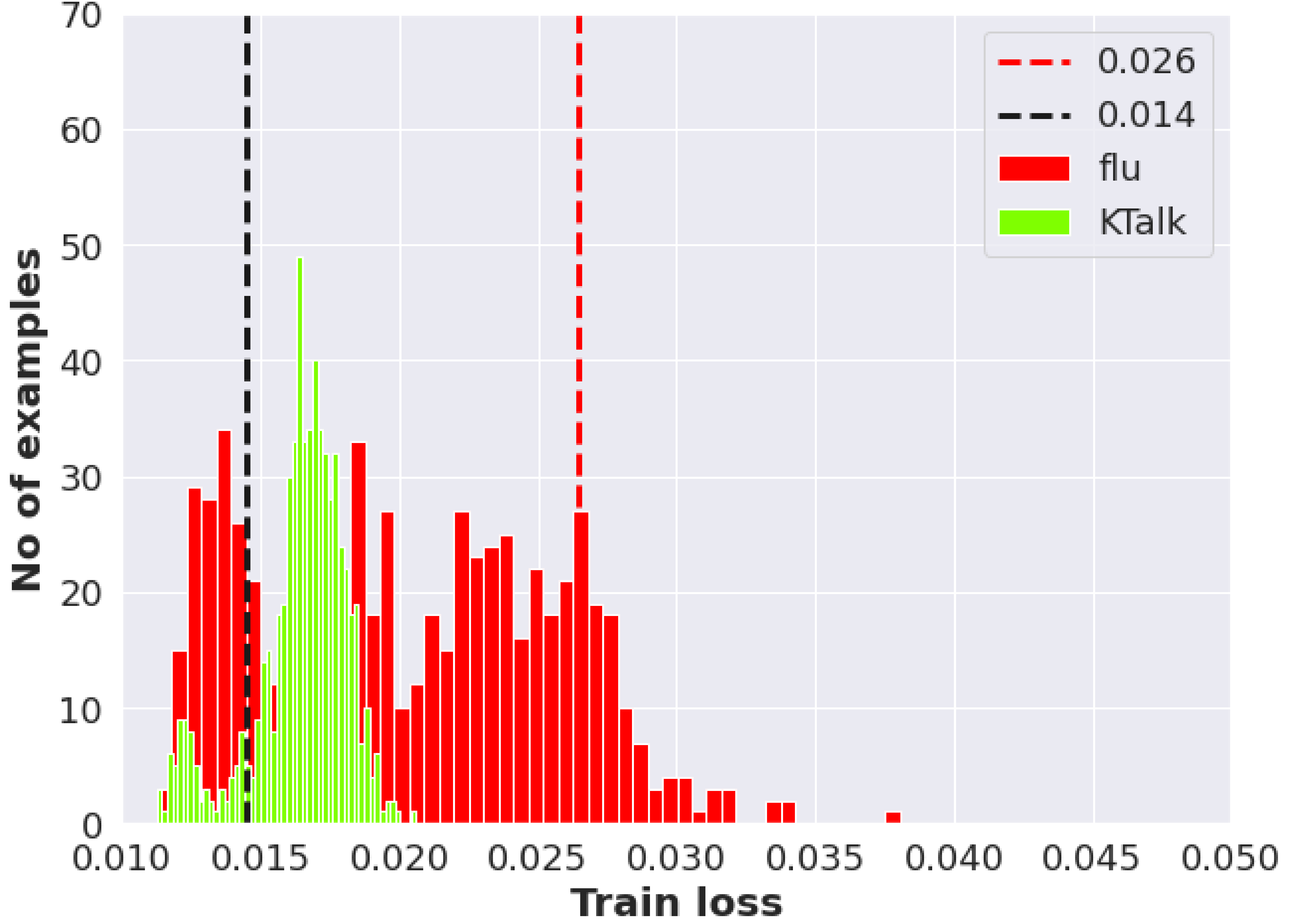}
    }
    \subfigure []{
        \label{subfig:Fig9C}
        \includegraphics[width=0.225\textwidth]{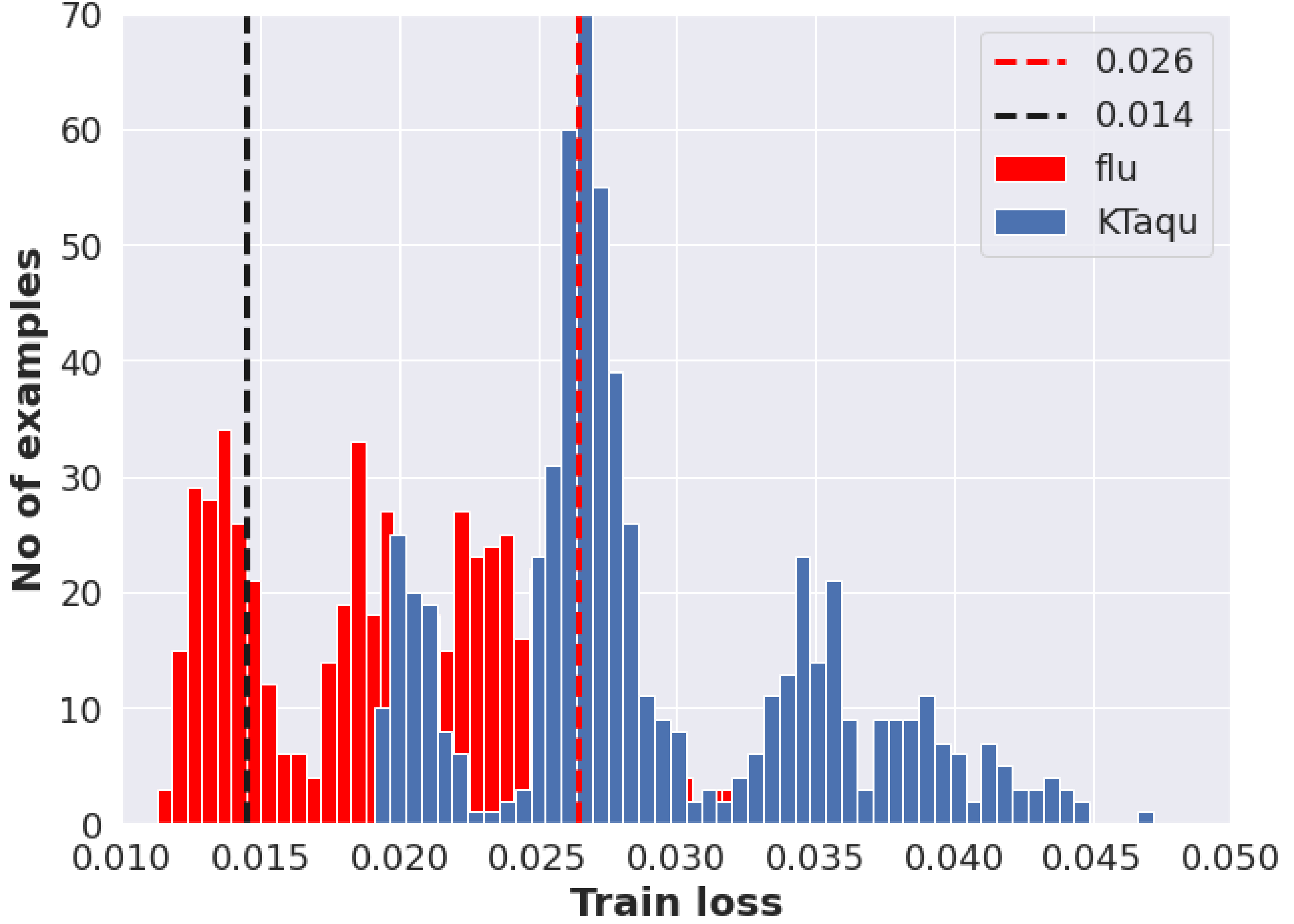}
    }
    \subfigure []{
        \label{subfig:Fig9D}
        \includegraphics[width=0.225\textwidth]{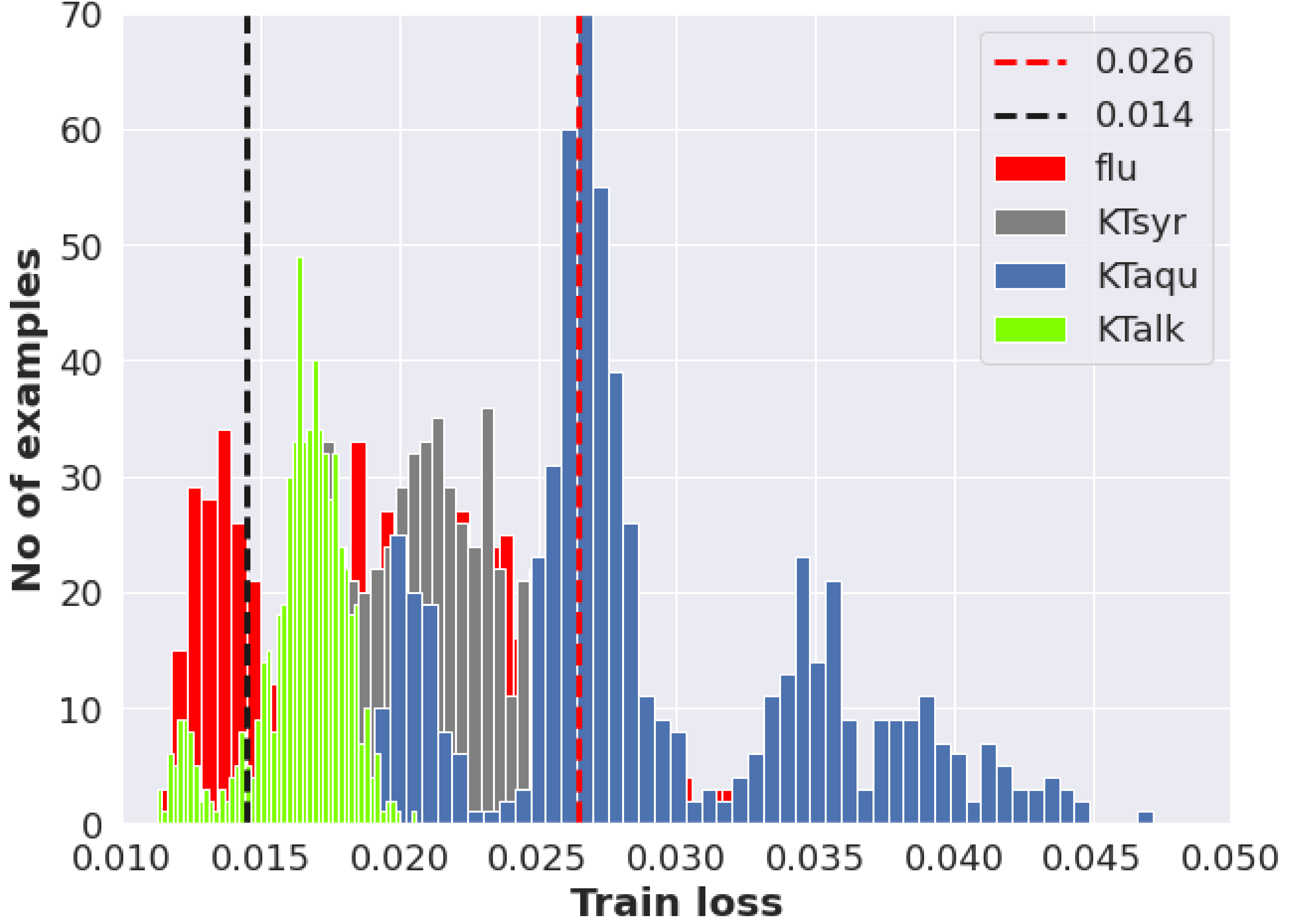}
    }
\caption{Histograms on anomaly tests were exhibited. The distribution of MSE loss from LFP features responded to the AD flu was illustrated along with each loss distribution of KT syrup (a), KT alkaloids (b), KT aqueous (c), and all forms of KT extracts (d). The lower and upper thresholds were specified for training loss at 0.026 and 0.014, labeled with black and red dash lines on each graph. Any samples dropped between these thresholds were indicated as similar to the LFP of the AD flu.}
\label{fig:Fig4}
\end{figure}

\begin{figure}
    \centering
    \subfigure []{
        \label{subfig:Fig6A}
        \includegraphics[width=0.225\textwidth]{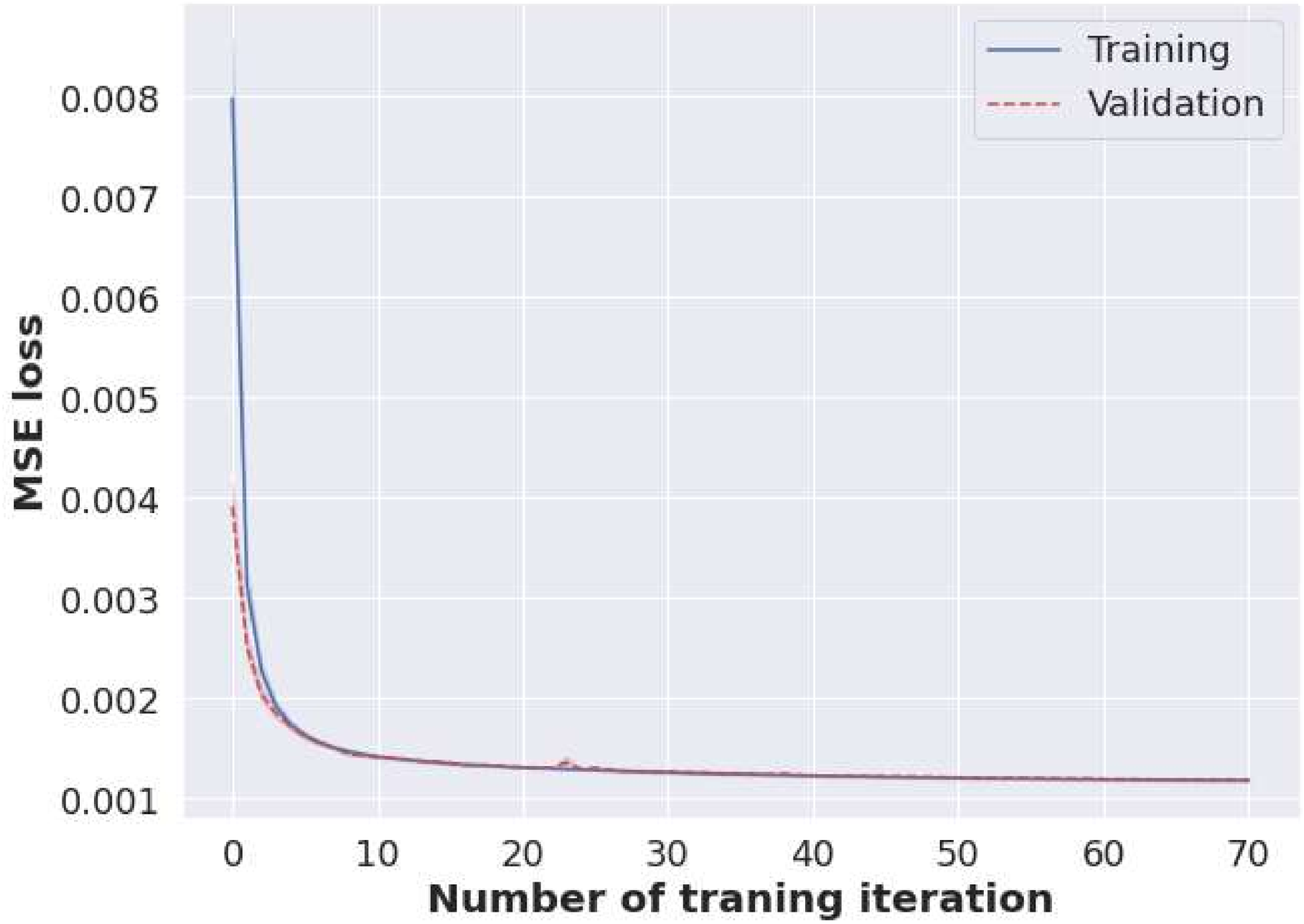} 
    }
    \subfigure []{
        \label{subfig:Fig6B}
        \includegraphics[width=0.225\textwidth]{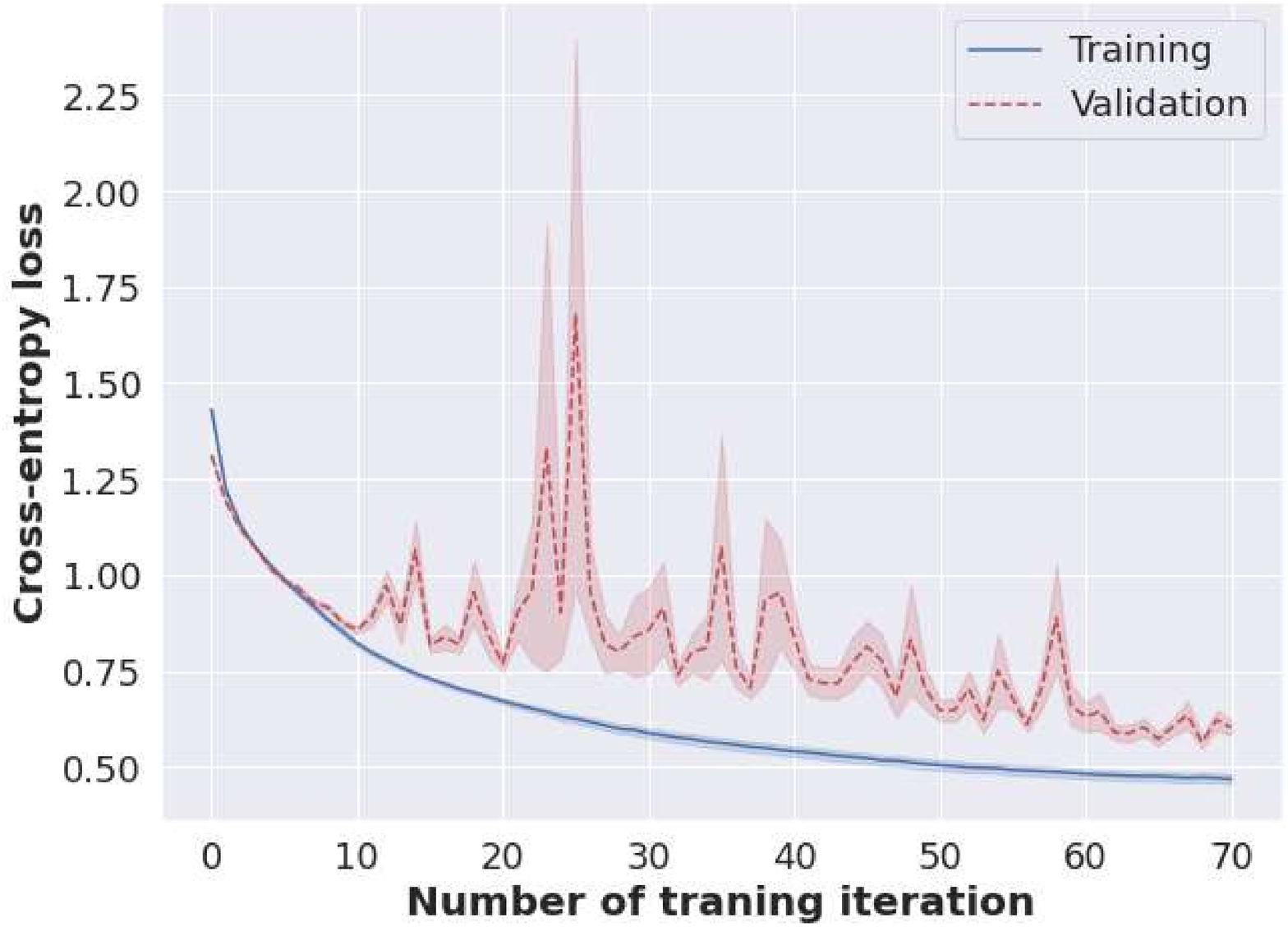}
    }
    
    \subfigure []{
        \label{subfig:Fig6C}
        \includegraphics[width=0.225\textwidth]{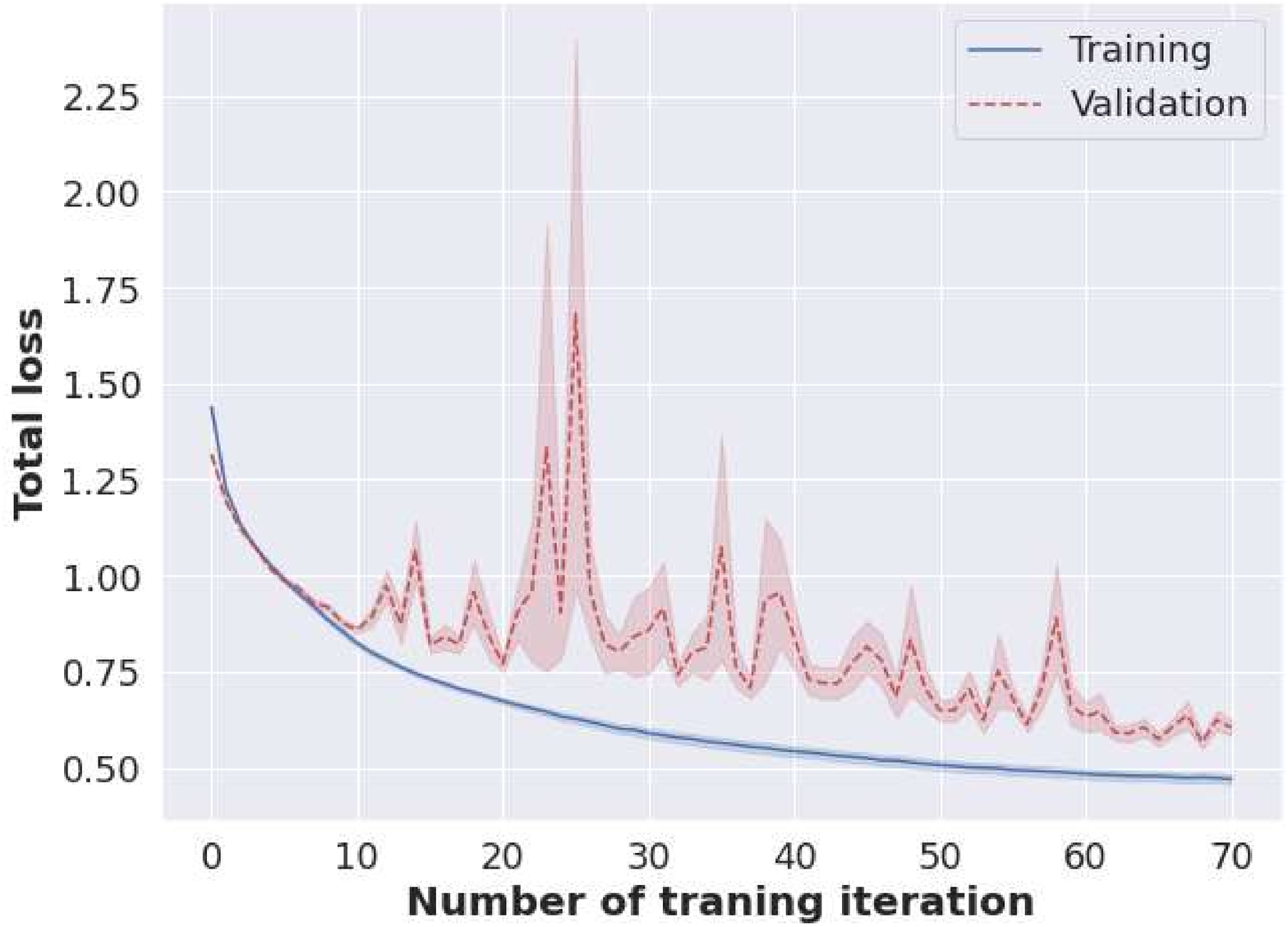}
    }
    \subfigure []{
        \label{subfig:Fig6D}
        \includegraphics[width=0.225\textwidth]{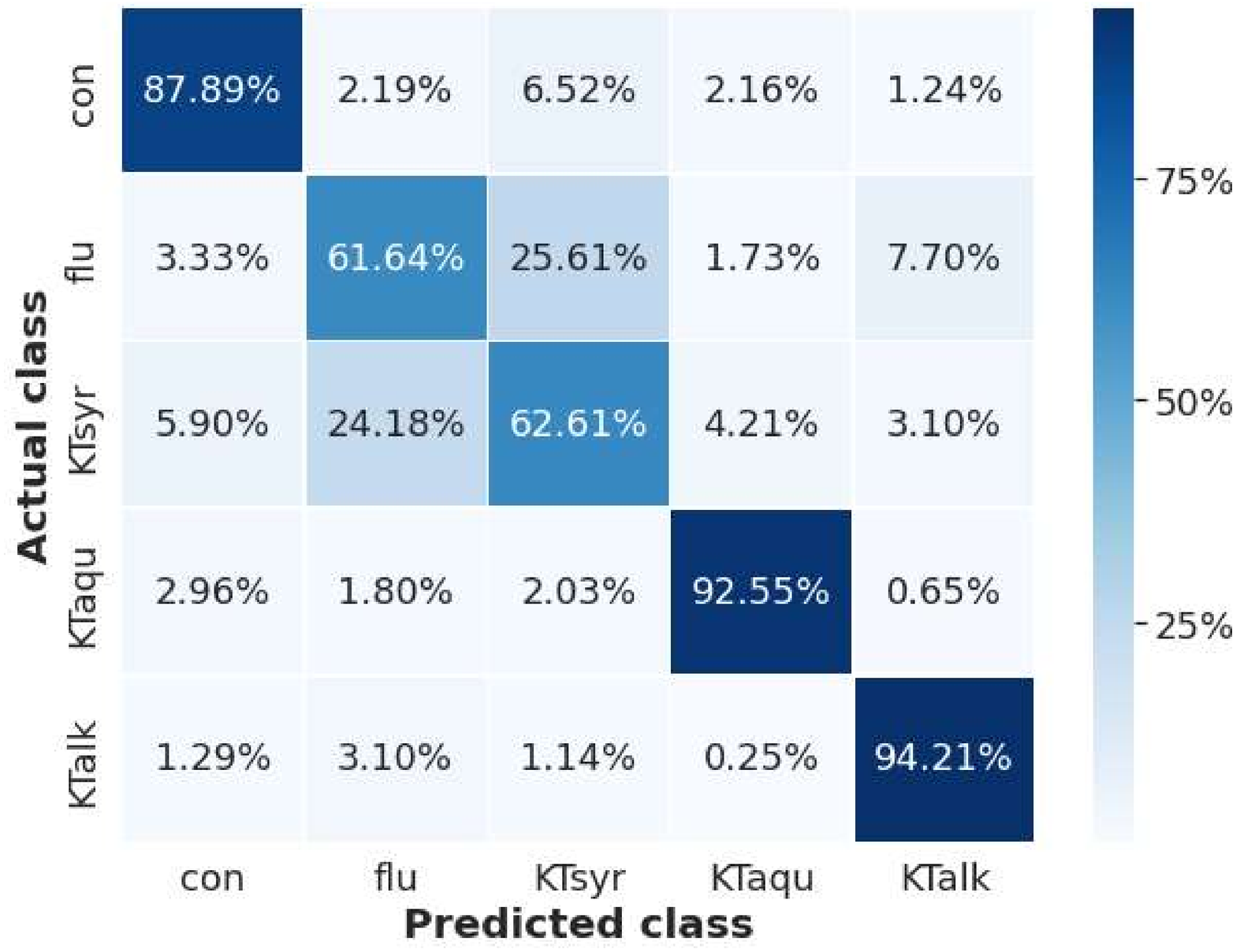}
    }

\caption{Training and validation losses were exhibited as the standard error mean (MSE) (a) and cross-entropy (b) loss. These two losses were summed to show as total loss (c). The confusion matrix of multi-LFP class recognition was shown (d). The con, flu, KTsyr, KTaqu, and KTalk represent the LFP features responded to the control, fluoxetine, KT syrup, KT aqueous, and KT alkaloids, respectively.}
\label{fig:Fig5}
\end{figure}

\begin{figure}
    \centering
    \subfigure []{
        \label{subfig:Fig7A}
        \includegraphics[width=0.225\textwidth]{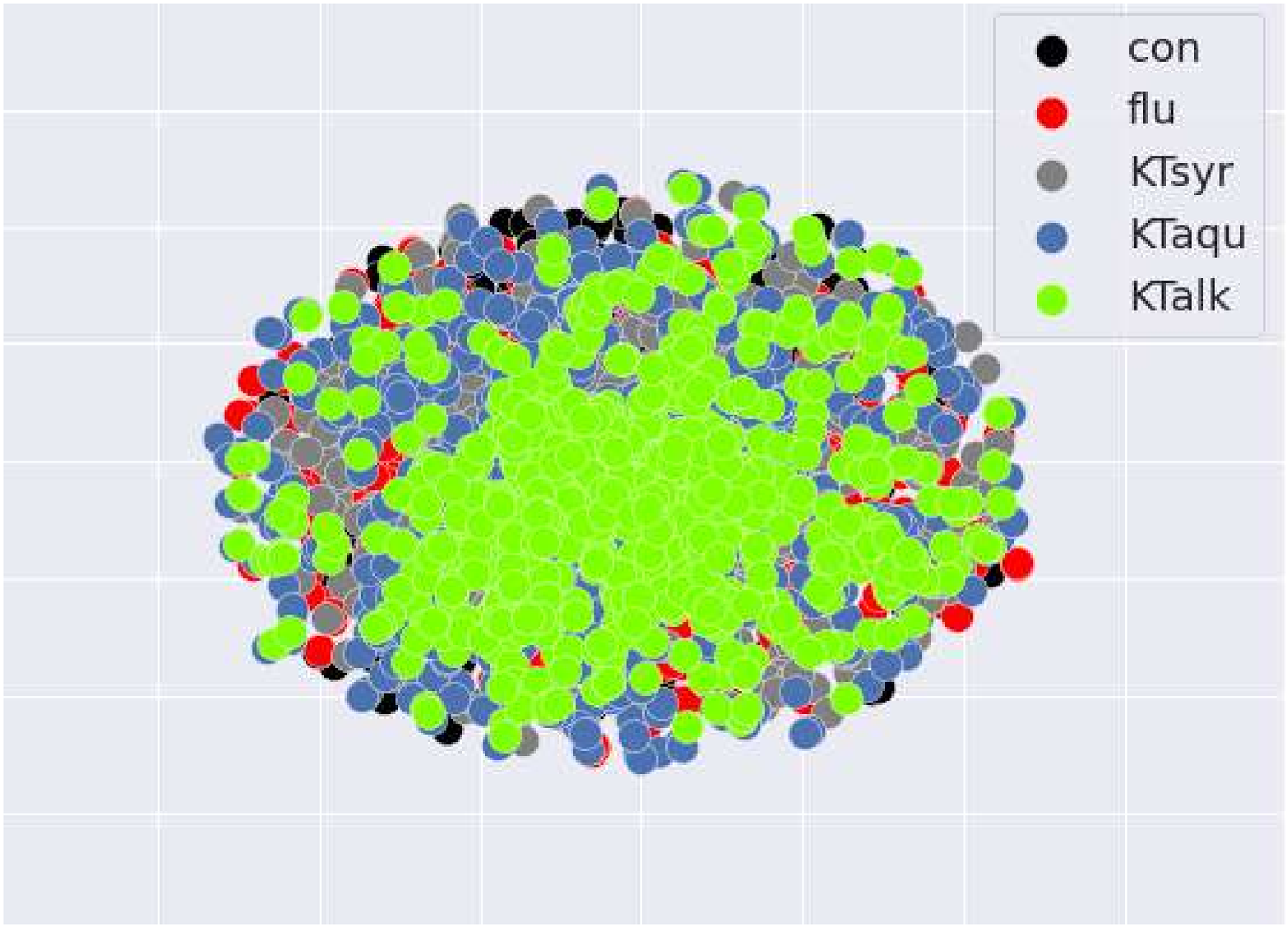} 
    }
    \subfigure []{
        \label{subfig:Fig7B}
        \includegraphics[width=0.225\textwidth]{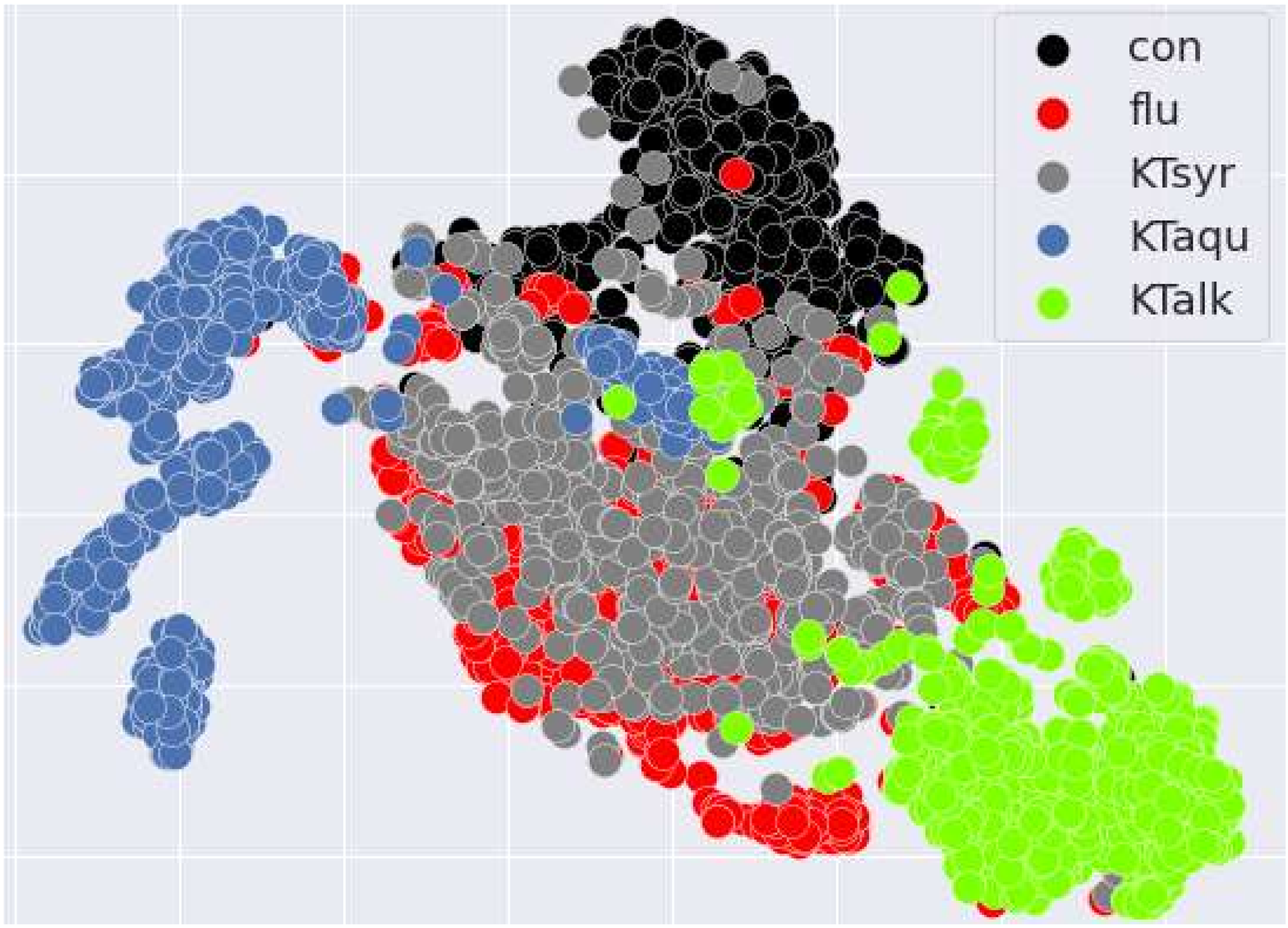}
    }    
\caption{Visualization of unlearned (raw spectral power data) (a) and learned (latent space) (b) LFP signal features by using 2-dimensional \textit{t}-SNE projection for the evaluation the LFP features responded to each substance: control (con), KT syrup (KTsyr), KT aqueous (KTaqu), KT alkaloids (KTalk), and fluoxetine (flu) in a subject-dependent setting}
\label{Fig6}
\end{figure}

\begin{figure}
    \centering
    \includegraphics[width=0.5\textwidth]{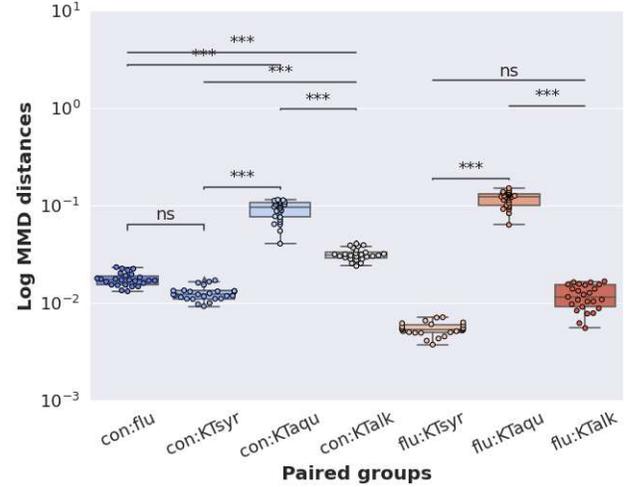}
    \caption{Boxplots of log MMD distances quantitatively assessed from \textit{t}-SNE projection were exhibited. For example, con: flu means the MMD distance measured between the embedding of LFP features responded to con and flu and so on. Asterisks indicate significant differences at ***: $p$ $<$ 0.001 performed by a non-parametric Kruskal-Wallis test with Bonferroni correction. The con, flu, KTsyr, KTaqu, and KTalk represent the LFP features responded to the control, fluoxetine, KT syrup, KT aqueous, and KT alkaloids, respectively.}
\label{fig:Fig7}
\end{figure}

\subsection{Experiment II: AE-Based Anomaly Detection (ANet)}\label{Experiment_II}
Experiment II demonstrated the performance of the proposed anomaly detector or ANet in assessing the similarity of the unseen LFP compared to LFP response to AD flu was the reference class or standard antidepressant substance in producing the drugs. The findings revealed that the number of LFP that responded to KT syrup appeared to distribute mainly in the normal zone indicating the high similarity to the standard drug or AD flu \autoref{fig:Fig4}(a), followed by KT alkaloids \autoref{fig:Fig4}(b), and KT aqueous \autoref{fig:Fig4}(c) extract, respectively. Quantitative data found that the normal detecting rate or defined as the similarity rate for this study, was 85.62 ± 0.29\%, 83.64 ± 0.32\%, and 40.69 ± 0.38\% in detecting LFP responses or testing sets from KT syrup, KT alkaloids, and KT aqueous extract, respectively, referred to LFP response of AD flu or the training sets.

\subsection{Experiment III:  Multi-Task AE}\label{Experiment_III}
To investigate an insightful optimization process of the multi-task AE, we assessed the alterations of training and validation losses in the multi-LFP class recognition during the training process. Based on the least iteration stopped by early stopping, \autoref{fig:Fig5}(a), (b), and (c) demonstrated training and validation loss while training for 70 epochs for MSE loss, CE loss, and total loss, respectively. The findings demonstrated that MSE loss reached a relatively stable level after data training for ten epochs. Meanwhile, high fluctuation patterns were in the CE and the total losses during the 20$^{th}$-40$^{th}$ epochs. However, validation loss converged to the training loss, suggesting that the model training had not been overfitted, confirming this study's well-trained model. All measures of the performance were applied together with the early stopping function. In summary, the proposed multi-task AE reached accuracy, and the F1-score of the model was 79.78 ± 0.39\% and 79.53 ± 0.00\%, respectively.

Furthermore, we also illustrated a confusion matrix of the predicted outputs. While the matrix revealed that almost all LFP samples that responded to the treated substances were primarily correctly classified, it was obvious that there were some misclassifications between the AD flu and KT syrup, as shown in \autoref{fig:Fig5}(d). Thus, LFP samples from both classes might have similar spectral power features originating from sharing mechanisms exerted on the CNS and the brain. The finding was consistent with Subsections \ref{Experiment_I} and \ref{Experiment_II}.

The 2-dimension of the \textit{t}-SNE projection visualized the probability distribution of reduced dimensional data at latent space. It was found that \textit{t}-SNE of baseline, processed from raw spectral power, diffused and mixed among classes of substances evenly to form one cluster, as shown in \autoref{Fig6}(a). Meanwhile, the characteristics of \textit{t}-SNE resulting from the learned features or represented latent vectors appeared to be formed for correct clusters. For example, consistency with the experiments mentioned earlier and results, among KT extracts, the KT syrup cluster shared the intersect clusters of the AD flu mostly, as depicted in \autoref{Fig6}(b).

Although discriminative LFP features of KT extract distributed with the AD flu overlappingly in the \textit{t}-SNE visualization, this result seems to provide only qualitative estimation. To get a concrete explanation, MMD distance, a value estimated from the \textit{t}-SNE projection-based-latent vector evaluation, was thus analyzed quantitatively. The results exposed that LFP features of the AD flu and KT syrup were the first groups showing the shortest MMD distance to control mice and were followed by KT alkaloids and KT aqueous, respectively. Moreover, features in LFP responses from the AD flu: KT syrup and AD flu: KT alkaloids were the first two paired classes that produced the shortest MMD distance. In contrast, the longest MMD distance was detected in the LFP features of AD flu to KT aqueous, as shown in \autoref{fig:Fig7}. This finding also brought us to the conclusive similarity of LFP responses obtained from the group of mice treated with KT syrup and AD flu.

\section{Discussion}
Here, we discussed the results of three experiments to identify the gap in the conventional approach to LFP responses comparison and analysis. Then we addressed the contributions of the proposed approaches, which were ANet and the multi-task AE, in bridging the identified gap. Finally, we explained the feasible impact of the finding from the application point of view on the behavioral brain responses to the alternative AD substances for drug formulation; Kratom or KT extracted solutions.

Alternation of spectral power amplitudes is directly reflected in the changes in NeuroBios activity \cite{cheaha2017effects}. In addition, various experiments have presented the feasibility of using spectral power characteristics of LFP as a biomarker for substance profile classification \cite{manor2021characterization}. Therefore, as demonstrated in Section \ref{Experiment_I}, we conducted spectral power analysis and statistical testing. LFP spectral features that responded to KT syrup and KT alkaloids have no significant difference in frequency responses overall brain regions when paired with LFP that responded to AD flu or standard substance for AD drug formulation. Therefore, we might infer that mice's brains were influenced by KT syrup and KT alkaloids, closely to AD flu.

To utilize and enhance the findings of the conventional analysis, we proposed the AE-based LFP feature extractor or ANet to automatically detect the similarity of LFP responses from the interested substances compared to the reference drug, which was AD flu in this study. According to Section \ref{Experiment_II}, the defined similarity rate of LFP features extracted by ANet corresponding to KT syrup gave us about two percent higher than those from KT alkaloids referred to LFP that responded to AD flu. These findings performed the feasible applications of ANet in pre-screening the alternative substances which affect the mice's brain similar to the standard or well-known substance in the future drug formulated study.

Then, we extended ANet to be the multi-task AE for the multi-LFP response recognition, a more sophisticated task than anomaly or similarity detector. The benefit of this task was enhancing, visualizing, and measuring the difference between LFP responses of the candidate substances, which could not be seen during the conventional spectral power analysis. Here, LFP responses of KT syrup and KT alkaloids were great examples. As reported in Section \ref{Experiment_I}, the conventional approach could not give us a convincing conclusion in comparing LFP responses from the group of mice drugged by KT syrup, KT alkaloids, and AD flu. On the other hand,  the proposed multi-task AE with qualitative and quantitative measures was quite promising in summarizing that the LFP responses of KT syrup were much closer to AD flu than those of KT alkaloids.

Similar to revealing the similarity of LFP responses to KT syrup and AD flu for the first time, our investigation can inspire people interested in applying neural network-based computational modeling as innovative tools to the field of brain behavioral studies responses to the alternative formulation of novel drugs.

\section{Conclusion}
This study proposed a neural network-based approach or ANet for comparing local field potential or LFP responses among the mice drugs with different substances. ANet automatically extracted the features from the unseen LFP responses and predicted the similarity rate to the reference or the standard responses used during the training process. Moreover, the extended ANet in the form of a multi-task AE presented the possibility of classifying multi-class LFP simultaneously. Here, we applied the proposed approaches to study the effect of Kratom, or KT extracted substances, compared to the standard AD drug via the mice's brain activities. Both qualitative and quantitative assessments were used throughout the studies. As far as we concern, It was the first study in the literature. The outcomes convinced KT extracted using syrup induced the high similarity of LFP responses to the standard antidepressant drug or AD fluoxetine. In conclusion, we might infer that KT syrup is the potential candidate substance for an antidepressant drug formulation. However, further confirmation of the clinical and molecular scaled studies remains a future challenge.

\bibliographystyle{IEEEtran}
\bibliography{references}

\end{document}